\documentclass[10pt,a4paper]{memoir}

\usepackage[british]{babel}
\usepackage[sort]{cite}

\usepackage{graphicx}
\usepackage{amsmath}
\usepackage{authblk}
\usepackage{url}
\usepackage[colorlinks=true,allcolors=blue,bookmarks=false]{hyperref}

\usepackage{csquotes}
\usepackage[usenames,dvipsnames]{color}
\usepackage{textcomp}
\usepackage{verbatim}

\usepackage{tjp-layout-article}

\newcommand{\tjpstitle}{First electrical White Rabbit absolute calibration inter-comparison}

\newcommand{\tjpsfooter}{\rule{\textwidth}{0.5pt}\par\small\tjpstitle\\\raisebox{-4pt}{\includegraphics[height=10pt]{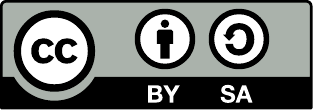}}\hspace{5pt}{\tiny This work is licensed under the \href{http://creativecommons.org/licenses/by-sa/4.0/}{Creative Commons Attribution-ShareAlike 4.0 International License.}}}

\makeevenhead{tjpheader}{}{}{}
\makeoddhead{tjpheader}{}{}{}
\makeevenfoot{tjpheader}{\thepage}{\tjpsfooter}{}
\makeoddfoot{tjpheader}{}{\tjpsfooter}{\thepage}

\graphicspath{{./Images/}}

\begin{document}

%%
%% Title page
%%
\title{\tjpstitle}

\author[1,*]{P.~P.~M.~Jansweijer}
\author[1]{N.~A.~D.~Boukadida}
\author[2]{K.~Hanhij\"{a}rvi}
\author[2]{A.~Wallin}
\author[3]{B.~Eglin}
\author[3]{E.~Laier English}

\affil[1]{Nikhef, Science Park 105, 1098 XG Amsterdam, The Netherlands}
\affil[2]{VTT Mikes, Tekniikantie 1, FIN-02150 Espoo, Finland}
\affil[3]{National Physical Laboratory, Hampton Road, Teddington, Middlesex, UK, TW11 0LW}
\affil[*]{Corresponding author: peterj@nikhef.nl}

\maketitle

{\vfill}

\section*{Copyright}
Copyright (C) 2022\\

\noindent{Nikhef (Peter P. M. Jansweijer, Nayib A. D. Boukadida),\\
VTT (Kalle Hanhij\"{a}rvi, Anders Wallin),\\
NPL (Belinda Eglin, Elizabeth Laier English)\\

\noindent{}\includegraphics{by-sa}

\noindent{}This work is licensed under the Creative Commons Attribution-ShareAlike 4.0 International License. To view a copy of this license, visit \href{http://creativecommons.org/licenses/by-sa/4.0/}{http://creativecommons.org/licenses/by-sa/4.0/} or send a letter to Creative Commons, PO Box 1866, Mountain View, CA 94042, USA.

%%
%% Main article text
%%
\newpage
\begin{center}
{\LARGE \tjpstitle}
\end{center}

\begin{abstract}
A time transfer link consisting of PTP White Rabbit (PTP-WR) devices can transfer time with sub-nanosecond accuracy.
Originally White Rabbit devices were calibrated as a set of two devices.
Progress in calibration makes individual absolute calibrated PTP-WR devices possible. 
This enables exchange of PTP-WR devices without the need for expensive in-situ end-to-end calibrations.  

Electrical absolute calibration is the basis of absolute calibration. 
It calibrates the time relationship between the internal timestamp and the external electrical time reference plane. 
In this paper we examine the electrical time transfer accuracy when a link is setup using electrical absolute calibrated PTP-WR devices calibrated by different laboratories.
\end{abstract}

\section{Introduction}

The open-source White Rabbit~\cite{WR} (PTP-WR) technology (integrated in the IEEE 1588-2019~\cite{IEEE1588-2019} Precision Time Protocol High Accuracy profile) provides sub-nanosecond accuracy and picosecond-level precision of synchronization over optical fibre, scalable to many nodes.

In June 2018, the White Rabbit for Industrial Timing Enhancement (WRITE~\cite{WRITE}) project was started.
One of the aims of WRITE is to further develop scalable calibration techniques for fibre propagation asymmetry, PTP-WR devices and SFPs (Small Form-factor Pluggable transceiver modules). Calibration of each individual component of a PTP-WR link is required for interoperability of PTP-WR equipment from different manufacturers and replacement in the event of equipment failures.
A system with calibrated components will in future lead to “plug-and-play” cost-effective time services to industry, replacing expensive in-situ end-to-end calibrations of each link.

Electrical absolute calibration is the basis of absolute calibration.
This paper focuses on the electrical absolute calibration of PTP-WR devices performed at different laboratories.
Independently calibrated PTP-WR devices were brought together for interoperability testing.
A PTP-WR time transfer link was set up.
The individual calibration constants were stored in each electrical absolute calibrated device.
The PPS residual between the devices was measured after the devices were brought in to full synchronization state.
Two main questions are posed in this paper: what is the accuracy of a time transfer link that uses PTP-WR devices
\begin{enumerate}
	\item which are calibrated at different laboratories?
	\item which are based on different hardware?
\end{enumerate}
The results are reported in this paper.

%\newpage
\section{Theory}
\label{sec:theory}

Time dissemination networks that are based on the Precision Time Protocol~\cite{IEEE1588-2019} timestamp the transmission and reception of messages in hardware.
Delays exist between the timestamp point in the interior of network components and their time reference planes.
For high accuracy time transfer these delays must be calibrated.

%This paper focuses on the time relations of the electrical time reference planes of PTP-WR devices as defined in~\cite{Abscal_ISPCS_2018}.

Electrical absolute calibration of PTP-WR devices is achieved by measuring the time relationship between the internal timestamps (t\textsubscript{1}, t\textsubscript{4p} and t\textsubscript{2p}, t\textsubscript{3}) and the external electrical time reference planes of each individual PTP-WR device (see Figure~\ref{fig:absolute_calibration_loopback}).
The calibration constants $\Delta$\textsubscript{TXcal} and $\Delta$\textsubscript{RXcal} include propagation delays due to hardware and gateware implementation.
The electrical time reference planes, as defined in~\cite{Abscal_ISPCS_2018}, consist of the connector that conveys the PPS inter-second boundary marker signal and the electrical interface to the electrical-optical/optical-electrical (EO/OE) converter that conveys the serial bits that encode the message timestamp point (MTP, defined in IEEE 1588-2019~\cite{IEEE1588-2019} paragraph 7.3.4.1).
In most cases an SFP module is used as an EO/OE converter, in which case its electrical time reference plane is the electrical SFP connector.

\begin{figure}[]
	\centering
	\includegraphics[width=0.70\textwidth]{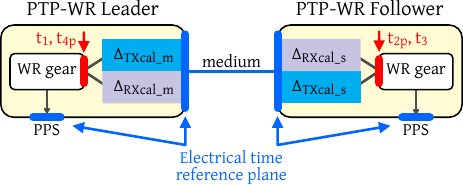}
	\caption{PTP-WR device electrical absolute calibration measures the time relationship between the internally generated timestamps t1, t4p and t2p, t3 (reference plane in red) and the external electrical time reference planes (blue) of each individual PTP-WR device (Leader and Follower). Timestamps t2p and t4p include fine delay phase information.}
	\label{fig:absolute_calibration_loopback}
\end{figure}

As described in~\cite{EE_abscal} cross correlation and interpolation techniques are used to determine time delay between two signals with high accuracy.
This can be done using a high speed real time Digital Storage Oscilloscope (DSO) that captures both PPS and serial bit stream as well as recording the corresponding White Rabbit timestamps.
However, PTP packets are created under software control and therefore the time between the rising edge of the PPS signal and the serial bits that encode the MTP varies and can be substantial, up to a few milliseconds.
This will require a very deep oscilloscope memory and finding a correlation peak will be computationally intensive.
A helper signal \enquote{abscal\_txts} was introduced in~\cite{Abscal_ISPCS_2018} to split the measurement into a non-deterministic time interval that can be measured with a Time Interval Counter (TIC) and a deterministic time interval that can be accurately measured with cross correlation and interpolation.
Unfortunately the two-step measurement decreases the accuracy of the calibration. 

Further details on the theory behind electrical absolute calibration of PTP-WR devices have been described in~\cite{Abscal_ISPCS_2018}, while the document \enquote{White Rabbit Electrical Absolute Calibration Procedure}~\cite{EE_abscal} describes how electrical absolute calibration is performed.

Two perfectly electrical absolute calibrated PTP-WR devices, connected via a medium without asymmetry, should have a zero time difference between their PPS signals.

%Two, perfectly electrical absolute calibrated PTP-WR devices that are connected by their electrical high speed link time reference planes should have a zero time difference between their PPS signals.

\section{Method}

Our purpose is to verify interoperability of devices of different type and calibration source.
Table~\ref{tab:el_abscal_devices} shows the different devices that were electrical absolute calibrated while Table~\ref{tab:el_abscal_source} shows the laboratories which were involved in calibration of the devices.
Devices are referred to by their name, serial-number and calibration source; for example SPEC\textsubscript{(VTT)} or SPEC7\textsubscript{(Sn01\_Nikhef)}.

\begin{table}[htbp]
	\centering
	\footnotesize
	\caption{\textbf{Electrical absolute calibrated devices}}
	\begin{tabular}{lcc}
		\hline
		    \textbf{Device name} & \textbf{abbreviation} & \textbf{reference} \\
		\hline
            Simple PCIe FMC carrier & SPEC & ~\cite{SPEC}  \\
            Simple PCIe FMC carrier 7 & SPEC7 & ~\cite{SPEC7} \\
%            KM3NeT Central Logic Board version 3 & CLBv3 & ~\cite{CLBv3} \\
		\hline
	\end{tabular}
    \label{tab:el_abscal_devices}
\end{table}

\begin{table}[htbp]
	\centering
	\footnotesize
	\caption{\textbf{Calibration source}}
	\begin{tabular}{lcc}
		\hline
            \textbf{Laboratory} & \textbf{abbreviation} & \textbf{reference} \\
 		\hline
            National Institute for Subatomic Physics & Nikhef & ~\cite{Nikhef} \\
            VTT-MIKES & VTT & \cite{VTT} \\
            National Physical Laboratory & NPL & ~\cite{NPL} \\
		\hline
	\end{tabular}
    \label{tab:el_abscal_source}
\end{table}

All devices are equipped with a 5 channel DIO card~\cite{DIO} plugged onto their FMC connector that outputs the PPS and abscal\_txts signals needed for electrical absolute calibration.
The plugged DIO card and plugged cables or adapters, to change from LEMO to SMA, are part of the electrical absolute calibration.
The definition of \enquote{device} spans this complete assembly as an indivisible entity.

\subsection{Determining $\Delta$\textsubscript{TXcal} and $\Delta$\textsubscript{RXcal}}
At Nikhef we used an Agilent 53230A Time Interval Counter (TIC) and a Keysight DSO-S-254A Digital Storage Oscilloscope (DSO) to perform the electrical absolute calibration as described in the document \enquote{White Rabbit Electrical Absolute Calibration Procedure}~\cite{EE_abscal}.
The 10MHz reference output of the TIC was used as reference for the DSO and as grandmaster clock for the PTP-WR device under calibration.
%White Rabbit operates in \enquote{mode abscal} during the calibration procedure.
%In mode abscal the PTP-WR device is phase locked to an external 10 MHz reference clock (equal to the behavior in grandmaster mode \enquote{mode gm}).
%The external 10 MHz reference clock re-lock might be of influence on the calibration results.

The thresholds for the TIC input channels affect the time interval measurement.
For each measurement the TIC input thresholds were set to 50\% of the input signal level.
The DSO time interval measurements are based on cross-correlation which is less susceptible in signal voltage levels.
Delay determination using cross-correlation is accurate under the condition that signal rise- and fall-times are equal.

%Although the overall method used for electrical absolute calibration is common and described in~\cite{EE_abscal}, details among the laboratories (i.e. their setups and procedures) may differ.
%Hence each calibration source has a dedicated section that describes their particular details.

Two different devices (SPEC and SPEC7, each with a 5 channel DIO card mounted as an indivisible entity) were electrical absolute calibrated.
%At first instance, the authors had reason to doubt the results listed in section \ref{sec:results}.
Calibration reproducibility was verified by performing several electrical absolute calibration sessions.
%(3 on SPEC\textsubscript{(VTT)}, 8 on SPEC\textsubscript{(Nikhef)}, 8 on each SPEC7).
%The mean values of calibration session results are taken as $\Delta$\textsubscript{TXcal} and $\Delta$\textsubscript{RXcal} for each device.
Electrical absolute calibration needs manual intervention during the measurement so calibration sessions were done on a table in the laboratory.

Devices were electrical absolute calibrated at different laboratories: Nikhef calibrated one SPEC and two SPEC7 devices, VTT calibrated one SPEC device and NPL calibrated two SPEC7 devices.
For inter-comparison, VTT sent their electrical absolute calibrated SPEC to Nikhef and Nikhef sent one SPEC7 to NPL.

\subsubsection{$\Delta$\textsubscript{TXcal} and $\Delta$\textsubscript{RXcal} at NPL}
NPL carried out the electrical absolute calibration as described above, but with different equipment (using a Tektronix TDS6124C DSO).
UTC(NPL) 10 MHz and PPS signals from a hydrogen maser were used as the grandmaster clock for the device under calibration.

\subsection{Inter-comparison of the calibrated devices}

\subsubsection{Inter-comparison at Nikhef}

The electrical absolute calibrated SPEC\textsubscript{(VTT)} was shipped from VTT in Espoo to Nikhef in Amsterdam where the devices were brought together and inter-comparison measurements were performed.

The devices were connected via a Direct Attach Cable with calibrated asymmetry.
%Propagation delay asymmetry between forward and backward channel of this cable is of influence on the measured residual PPS skew.
%Direct Attach Cable measurement~\cite{DAC_report} shows that the asymmetry is in the orr of 5 ps which contributes to a maximum of 2.5 ps PPS skew error.
The calibration constants $\Delta$\textsubscript{TXcal} and $\Delta$\textsubscript{RXcal} were stored in the corresponding calibrated devices.
One device was put in grandmaster mode that locks to a Keysight 33600A waveform generator which generated 10 MHz and PPS.
%\footnote{
%	When the Keysight 33600A waveform generator is locked to the 10MHz reference output of the Agilent 53230A TIC, the waveform generator PPS output and the internal TIC clock source interfere and produce a periodic (250 second) saw-tooth like effect which offsets the actual TIC measurement ($\approx$30 ps peak-peak).
%	This effect does not occur on the 10MHz generated output (i.e. it has a fixed phase relation with respect to the 10 MHz reference input).
%	During calibration PPS is only necessary to obtain a lock in \enquote{mode abscal}.
%	Therefore, calibration and PPS residual TIC measurements are not distorted by the effect described above.
%	However, for TIC skew measurements the waveform generator must not be locked to the 10MHz output of the TIC to enable true randomization of the measurements without the saw-tooth like disturbance.
%}.

%The Agilent 53230A Time Interval Counter (TIC) was setup to measure the time interval between Channel-1 \enquote{start} and Channel-2 \enquote{stop}.

SMA cables were kept connected to the inputs of the TIC.
The SMA connectors at the far end of the cables were the TIC reference planes for which skew was measured and taken into account.
The TIC reference planes were connected to the PPS reference planes of both PTP-WR devices.
In an automated set up, the PTP-WR link was repeatedly brought into full synchronization (PTP-WR nomenclature \enquote{TRACK\_PHASE}) in order to randomize link restart effects.
After each link restart, the mean value of 600 PPS residual measurements was recorded with the TIC while the link was in \enquote{TRACK\_PHASE}.
In this way one measurement takes at least 10 minutes so overnight this resulted in 80 link-restart measurements.

\subsubsection{Inter-comparison at NPL}

The SPEC7\textsubscript{(Sn07\_Nikhef)}, that was electrical absolute calibrated at Nikhef, was shipped to NPL in Teddington where the devices were brought together and inter-comparison measurements were performed.

The calibration constants $\Delta$\textsubscript{TXcal} and $\Delta$\textsubscript{RXcal} were stored in the corresponding calibrated devices. One device was put in grandmaster mode and locked to UTC(NPL) 10 MHz and PPS signals, generated from a hydrogen maser, then connected to the other via a single mode fibre optic cable and two SFPs. Both devices outputted PPS signals, fed to the Keysight 53230A TIC through SMA cables. The skew of the measurement setup was measured from the ends of the cables furthest from the TIC. 

The NPL setup was not automated, and an unknown asymmetry was introduced by the SFPs and fibre.
The measurements were performed in sets to compensate for this asymmetry.
The PPS residual was determined to be the mean value of the measurements before and after switching the SFPs and fibre.
600 PPS residual measurements were taken, the switch took place, and then another 600 were taken.
This was then repeated as time allowed.
\section{Systematic effects and uncertainties}
Table~\ref{tab:error_sources} is a summary of the error sources and their significance.
\begin{table}[htbp]
	\centering
	\footnotesize
	\caption{\textbf{Type B measurement uncertainties.}}
	\begin{tabular}{lrrr}
		\hline
		type B uncertainty & mean (ps) & $u_{j}$ (ps)\\
		\hline
		TIC temperature dependency & 0.0 & 12.5\\
		DSO temperature dependency & 0.0 & 0.8\\
        TIC reproducibility & 0.0 & 4.0\\
        DSO reproducibility & 0.0 & 0.7\\
        Cable Manipulation & 0.0 & 7.5\\
		Direct Attach Cable Asymmetry & 0.0 & 1.3\\
		\hline
	\end{tabular}\label{tab:error_sources}
\end{table}

Calibration accuracy depends on the stability of the Agilent 53230A Time Interval Counter and the Keysight DSO-S-254A Digital Storage Oscilloscope.
Measurements are averaged over 600 samples to average the single-shot accuracy.

Both TIC and DSO were subjected to a test in the climate chamber while cycling temperature over the range from 15 to 35\textcelsius{}.
Time interval measurements showed a +5 ps/\textcelsius{} relation.
The DSO showed a -0.3 ps/\textcelsius{} relation.

Calibration of PTP-WR devices requires manual intervention and had to be performed outside the climate chamber in a laboratory where room temperature was 22.5\textcelsius{} $\pm$2.5\textcelsius{}.
As described in ~\cite{EE_abscal}, calibration depends on the sum of TIC and DSO measurements.
When the calibration measurements are performed at the same laboratory temperature (i.e. shortly after one-another) then the positive and negative temperature coefficients of TIC and DSO respectively partly cancel out.
However, to account for the worst case scenario, we take them as independent; the temperature dependency of the TIC is taken to be $\pm$12.5 ps while the dependency for the DSO is taken to be $\pm$0.8 ps.

Both TIC and DSO measurements were tested for reproducibility at constant 23.5\textcelsius{} over 24 hours to be within $\pm$4 ps and $\pm$0.7 ps respectively.

SMA cable manipulation resulted in $\pm$7.5 ps uncertainty contribution.

%To examine calibration reproducibility several electrical absolute calibration sessions were performed (3 on SPEC\textsubscript{(VTT)}, 8 on SPEC\textsubscript{(Nikhef)}, 8 on each SPEC7).

Propagation delay asymmetry between forward and backward channels of the Direct Attach Cable, that is used to connect both serial high speed electrical time reference planes of the calibrated PTP-WR devices, is of influence on the measured PPS residual.
Direct Attach Cable measurement~\cite{DAC_report} shows that the asymmetry is 5 ps which contributes to a maximum of 2.5 ps PPS residual error (i.e. $\pm$1.3 ps).

The combined contribution of Table~\ref{tab:error_sources} Type B measurement uncertainties is 15.2 ps.

\section{Experimental results}
\label{sec:results}
\subsection{Electrical absolute calibration constants $\Delta$\textsubscript{TXcal} and $\Delta$\textsubscript{RXcal}}
Table~\ref{tab:absacal_constants} shows the measurement results of the electrical absolute calibrations.

\begin{table}[htbp]
	%\begin{table}[H]
	\centering
	\footnotesize
	%\scriptsize
	\caption{\textbf{Electrical absolute calibration constants $\Delta$\textsubscript{TXcal}, $\Delta$\textsubscript{RXcal} and their uncertainty $\sigma$ in ps.
	Serial numbers of SPEC, SPEC7 and DIO cards are listed.
	Bitfiles used: SPEC\textsubscript{VTT} $\rightarrow$ spec-dio\_nic-v3.0.bit; SPEC\textsubscript{Nikhef} $\rightarrow$ spec\_wr\_ref\_top\_elf.mcs (181121); all SPEC7's $\rightarrow$ spec7\_ref\_top\_z035\_210121\_1600\_elf.bit}}
	
	%\resizebox{\columnwidth}{!}{%
	\begin{tabular}{ lcccccc }
		\hline
		       & SPEC & SPEC   & SPEC7  & SPEC7  & SPEC7  & SPEC7\\
		\hline
		 Laboratory & VTT  & Nikhef & Nikhef & Nikhef & NPL & NPL\\
		\hline
		$\Delta$\textsubscript{TXcal} & 29394   & 30054   & 64938   & 65168   & 65626	& 65467\\
		$\Delta$\textsubscript{RXcal} & -297789 & -297853 & -248558 & -248137 & -244765	& -244887\\
		$\sigma$                      & 66.1    & 62.2    & 30.5    & 29.2    & 42.1	& 39.4\\
		\hline
		%$n$ & 1 & 1 & 4 & 4 & ?\\
		%\hline
		Sn & 7S-O-S9 0001 & 0183v4 & Sn\_01 & Sn\_07 & Sn\_10 & Sn\_11\\
		DIO\_Sn & 7S-2.0-S7.003 & 0038v2 & 0057v2 & 0058v2 & S12.206 & S12.205\\
		\hline
	\end{tabular}
	
	\label{tab:absacal_constants}
\end{table}

The uncertainties $\sigma$ for the SPEC listed in Table~\ref{tab:absacal_constants} are higher than for SPEC7.
This was expected since effort was put into the design of SPEC7 to improve the WR hardware (WRITE task 3.1 \cite{WRITE}).
%The quality and stability of the SPEC7 Voltage Controlled Crystal Oscillator (VCXO) reduces the need for the White Rabbit PLL to tune the VCXO~\cite{WR_Clocks} which significantly reduces jitter on PPS$\rightarrow$abscal\_txts and PPS$\rightarrow$PPS measurements.
%The next section describes PPS residual measurements.
%As can be seen in table \ref{tab:absacal_constants} different types of PTP-WR devices have different electrical absolute calibration uncertainties ($\sigma$).
Table \ref{tab:accumulated_uncertainties} lists the expected accumulated uncertainties when a White Rabbit link is set up between different types of PTP-WR devices.
Table \ref{tab:accumulated_uncertainties} includes the Type B measurement uncertainties of Table \ref{tab:error_sources}.

\begin{table}[htbp]
	\centering
	\footnotesize
	\caption{\textbf{Expected accumulated uncertainties for PPS residual measurements with different combinations of PTP-WR devices.}}
	\begin{tabular}{cc}
		\hline
		PTP-WR devices & $\sigma$ (ps)\\
		\hline
		SPEC7$\leftrightarrow$SPEC7 & 45\\
		SPEC7$\leftrightarrow$SPEC  & 75\\
		SPEC$\leftrightarrow$SPEC   & 93\\
		\hline
	\end{tabular}\label{tab:accumulated_uncertainties}
\end{table}

\subsection{Inter-comparison results of the calibrated devices}

Time transfer links were setup for 7 combinations (A-G) of the electrical absolute calibrated PTP-WR devices listed in Table \ref{tab:absacal_constants}.
Table~\ref{tab:intercomp} shows the mean values of \enquote{n} link restart PPS residual measurements\footnote{Where n = 77 or 78 a few outliers due to misbehaving hardware/gateware were removed.}.
Both link directions are listed, i.e. for each the combination PTP-WR device grandmaster/follower roles are swapped.

In accordance with the definition of IEEE1588 (see~\cite{IEEE1588-2019} figure N.3) a PPS residual is positive if the transition at the follower PPS output occurs later than the transition at the leader PPS output.
When determining the mean PPS residual ($\mu$) the sign of the second measurement should be inverted.

The PPS residual measurements for combinations A to D shown in Table~\ref{tab:intercomp} were performed in a laboratory at room temperature (22.5\textcelsius{} $\pm$2.5\textcelsius{}).
The stabilities of these measurements, even with some temperature variation, were satisfactory ($\sigma$=3.6 ps, see also figure \ref{fig:meas_210329}).
The gateware used for SPEC7's for combinations A to D is based on (currently not yet officially released) wr-cores~\cite{wr-cores} version 5 which implements improved Gigabit Transceiver SerDes (GTX) determinism.
% which ensures a lock at bitslide=0.

\begin{figure}[h]
	\centering
	\includegraphics[width=0.7\textwidth]{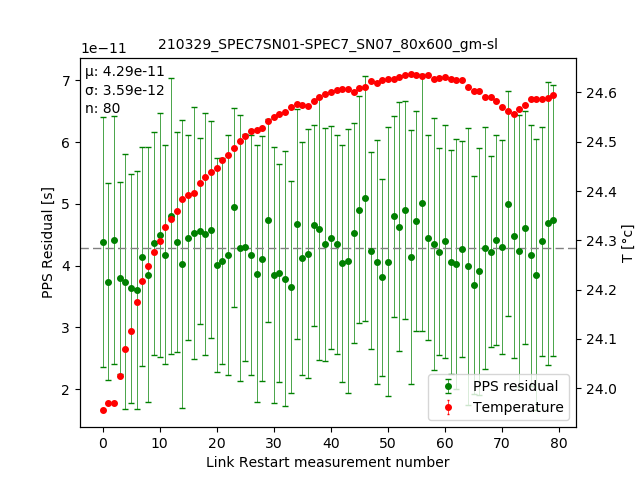}
	\caption{Link restart PPS residual measurements performed in the laboratory outside the climate chamber. SPEC7\textsubscript{Sn01\_Nikhef} (grandmaster) to SPEC7\textsubscript{Sn07\_Nikhef} (follower).
	SPEC7 gateware uses wr-cores version 5.}
	\label{fig:meas_210329}
\end{figure}

The combinations E-G each involve SPEC cards which are  based on wr-cores version 4.2.
At startup, the SPEC local oscillator locks at a random position w.r.t. the link serial bitstream and phase is compensated for by taking bitslide into account (see \enquote{Semi-static latency} M.4.2 of \cite{IEEE1588-2019}).
Unfortunately the various lock positions with respect to the serial bitstream are not precisely spaced at 800 ps unit intervals, which for the SPEC, leads to mismatches in the order of 30 ps.
This phenomenon can be seen in Figure \ref{fig:meas_210524} where \enquote{bands} can be distinguished around 90, 120 and 150 ps.
It is due to this behavior that link restarts are necessary for finding a proper mean value by randomizing enough measurements.

\begin{figure}[]
	\centering
	\includegraphics[width=0.7\textwidth]{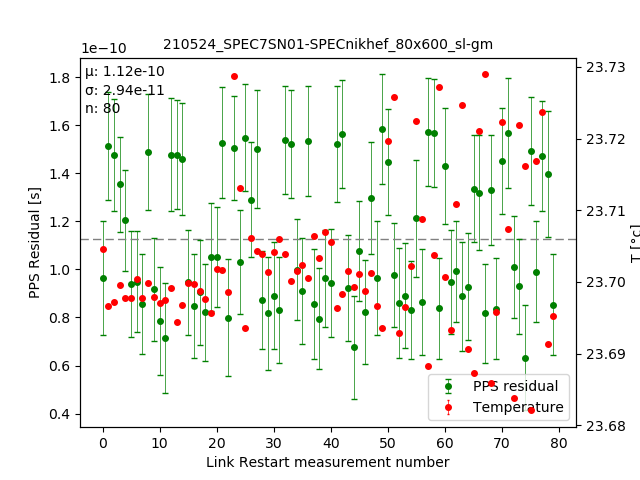}
	\caption{Link restart PPS residual measurements. SPEC\textsubscript{Nikhef} (grandmaster) to SPEC7\textsubscript{Sn01\_Nikhef} (follower).
	\enquote{Bands} are visible due to wr-cores v4.2 bitslide selection.}
	\label{fig:meas_210524}
\end{figure}

To avoid mixing link restart uncertainties with uncertainties due to temperature dependency we placed combinations E-G in a climate chamber to minimize temperature influence.
Unfortunately SPEC7\textsubscript{Sn07\_Nikhef} had already been shipped to NPL so re-doing measurement A in the climate chamber was no longer an option.

\begin{table}[h!]
	\centering
	\footnotesize
	%\scriptsize
	\caption{\textbf{Inter-comparison results for different combinations A to G of PTP-WR devices. Combinations A-D were measured in a laboratory at room temperature: 22.5\textcelsius{} ($\pm$2.5\textcelsius{}) while combinations E-G were measured in a climate chamber at 23.5\textcelsius{} ($\pm$0.1\textcelsius{}).}}
	%\resizebox{\columnwidth}{!}{%
	\begin{tabular}{ l l c  c  c  c }
		\cline{2-6}
		\multicolumn{1}{l|}{ } & {} & PPS residual & n & $\sigma$ & figure \\
		\multicolumn{1}{l|}{ } & {grandmaster $\rightarrow$ follower} & (ps) &   & (ps) & \\
		\hline
		\multicolumn{1}{l|}{ } & {SPEC7\textsubscript{Sn01\_Nikhef}$\rightarrow$SPEC7\textsubscript{Sn07\_Nikhef}} & 43         & 80 & 3.6 & \ref{fig:meas_210329}\\
		\multicolumn{1}{l|}{A} & {SPEC7\textsubscript{Sn07\_Nikhef}$\rightarrow$SPEC7\textsubscript{Sn01\_Nikhef}} & -64 & 77 & 3.4 & \\
		\cline{2-6}
		\multicolumn{1}{l|}{ } & {$\mu$, $\Delta$ (ps)}                                                            & 54, 21     &    &     & \\
		\hline
		\hline
		\multicolumn{1}{l|}{ } & {SPEC7\textsubscript{Sn07\_Nikhef}$\rightarrow$SPEC7\textsubscript{Sn10\_NPL}}    & 536          & 6  & 19.1   & \\
		\multicolumn{1}{l|}{B} & {SPEC7\textsubscript{Sn10\_NPL}$\rightarrow$SPEC7\textsubscript{Sn07\_Nikhef}}    & -524  & 8  & 16.4   &\\
		\cline{2-6}
		\multicolumn{1}{l|}{ } & {$\mu$, $\Delta$ (ps)}                                                            & 530, 12      &    &     &\\
		\hline
		\hline
		\multicolumn{1}{l|}{ } & SPEC7\textsubscript{Sn07\_Nikhef}$\rightarrow$SPEC7\textsubscript{Sn11\_NPL}   & 491         & 6 & 23.5 & \\
		\multicolumn{1}{l|}{C} & {SPEC7\textsubscript{Sn11\_NPL}$\rightarrow$SPEC7\textsubscript{Sn07\_Nikhef}} & -464 & 6 & 29.0 & \\
		\cline{2-6}
		\multicolumn{1}{l|}{ } & {$\mu$, $\Delta$ (ps)}                                                  & 478, 27    &    &      & \\
		\hline
		\hline
		\multicolumn{1}{l|}{ } & SPEC7\textsubscript{Sn10\_NPL}$\rightarrow$SPEC7\textsubscript{Sn11\_NPL}   & -37       & 6 & 15.4 & \\
		\multicolumn{1}{l|}{D} & {SPEC7\textsubscript{Sn11\_NPL}$\rightarrow$SPEC7\textsubscript{Sn10\_NPL}} & 13 & 6 & 21.4 & \\
		\cline{2-6}
		\multicolumn{1}{l|}{ } & {$\mu$, $\Delta$ (ps)}                                                  & 25, 24   &    &      & \\
		\hline
		\hline
		\multicolumn{1}{l|}{ } & {SPEC\textsubscript{Nikhef}$\rightarrow$SPEC\textsubscript{VTT}} & 96          & 80 & 37.6 & \\
		\multicolumn{1}{l|}{E} & {SPEC\textsubscript{VTT}$\rightarrow$SPEC\textsubscript{Nikhef}} & -119 & 80 & 38.4 & \\
		\cline{2-6}
		\multicolumn{1}{l|}{ } & {$\mu$, $\Delta$ (ps)}                                           & 107, 23     &    &      & \\
		\hline
		\hline
		\multicolumn{1}{l|}{ } & {SPEC7\textsubscript{Sn01\_Nikhef}$\rightarrow$SPEC\textsubscript{Nikhef}} & 162         & 80 & 32.2 & \\
		\multicolumn{1}{l|}{F} & {SPEC\textsubscript{Nikhef}$\rightarrow$SPEC7\textsubscript{Sn01\_Nikhef}} & -112 & 80 & 29.4 & \ref{fig:meas_210524}\\
		\cline{2-6}
		\multicolumn{1}{l|}{ } & {$\mu$, $\Delta$ (ps)}                                                     & 137, 50     &    &      & \\
		\hline
		\hline
		\multicolumn{1}{l|}{ } & SPEC7\textsubscript{Sn01\_Nikhef}$\rightarrow$SPEC\textsubscript{VTT}   & 241         & 77 & 24.0 & \\
		\multicolumn{1}{l|}{G} & {SPEC\textsubscript{VTT}$\rightarrow$SPEC7\textsubscript{Sn01\_Nikhef}} & -208 & 78 & 19.1 & \\
		\cline{2-6}
		\multicolumn{1}{l|}{ } & {$\mu$, $\Delta$ (ps)}                                                  & 224, 33     &    &      & \\
		\hline
		
	\end{tabular}
	\label{tab:intercomp}
\end{table}

Ideally, the PPS residual should be zero in both link directions.
%The Difference $\Delta$ is listed in the table \ref{tab:intercomp} as well as the mean PPS residual $\mu$.
When swapping grandmaster/follower roles of the PTP-WR devices there is a PPS residual difference $\Delta$ due to link asymmetry which causes a deviation from the mean PPS residual value $\mu$, both listed in Table \ref{tab:intercomp}.

%The $\Delta$ results in table \ref{tab:intercomp} show that swapping grandmaster/follower roles of the PTP-WR devices has an effect on the PPS residual.
%Moreover, the PPS residual mean value $\mu$ is bigger than expected.

Combinations \enquote{A} and \enquote{D} both use equal type PTP-WR devices (SPEC7).
For combination \enquote{A}, both devices were electrical absolute calibrated at Nikhef while for combination \enquote{D}, both devices were calibrated at NPL.
The resulting $\mu$=25 ps for combination \enquote{D} is within the 45 ps expected accumulated uncertainty (see Table \ref{tab:accumulated_uncertainties}) for this combination, while $\mu$=54 ps for combination \enquote{A} is outside expected accumulated uncertainty.

Combinations \enquote{B} and \enquote{C} both use equal type PTP-WR devices (SPEC7) which were calibrated at different laboratories (Nikhef and NPL).
Consistent large offsets ($\mu\approx{504}$ ps) exist between the devices that were calibrated at these laboratories.

Combination \enquote{E} compares equal type PTP-WR devices (SPEC) that were electrical absolute calibrated at different laboratories (Nikhef and VTT).
The mean PPS residual of $\mu$=107 ps is also outside the 93 ps expected accumulated uncertainty listed for this combination (see Table \ref{tab:accumulated_uncertainties}).

Combination \enquote{F} compares different types of PTP-WR devices (SPEC and SPEC7) both electrical absolute calibrated at Nikhef.
The offsets 162 ps, -112 ps ($\mu$=137) and difference ($\Delta$=50 ps) are significant.

Combination \enquote{G} compares different types of PTP-WR devices (SPEC and SPEC7) both electrical absolute calibrated at different laboratories (Nikhef and VTT).
The results show large offsets ($\mu$=224), outside the 75 ps expected accumulated uncertainty (Table \ref{tab:accumulated_uncertainties}).

The significant offsets and link direction differences of the PPS residual measurements listed in Table \ref{tab:intercomp} encouraged us to re-verify and redo the electrical absolute calibration multiple times and using different equipment to re-assure that proper $\Delta$\textsubscript{TXcal} and $\Delta$\textsubscript{RXcal} values were obtained.

%\newpage
\section{Discussion}
\label{sec:discussion}

The results of the measurements are not what the authors had anticipated.
Except for combination \enquote{D} the large offsets are clearly outside the expected uncertainties.

The mismatch for combinations \enquote{B}, \enquote{C} and \enquote{E} could be explained by the fact that equal type PTP-WR devices were electrical absolute calibrated at different laboratories, using different calibration setups.
However, the result of combination \enquote{A} (i.e. an offset for equal type PTP-WR devices calibrated at the same laboratory) shows that the above statement is not fully justified.

The large offsets for combinations \enquote{F} and \enquote{G} seem to be due to differences in hardware.
SPEC and SPEC7 are build around a Xilinx Spartan-6 and Zynq-7000 FPGA respectively.
Electrical absolute calibration determines the time relation between the internal PTP time stamps and the external electrical time reference plane and therefore, any hardware/gateware differences were expected to be covered.

The authors had various discussions with experts to find possible causes for any hardware dependencies.
One of the suggestions is that the CMOS processes for SPEC Spartan-6 (45 nm) and SPEC7 Zynq-7000 (28 nm) differ, hence the electrical signals on SPEC7 have faster rise and fall times.
The 20-80\% rise-/fall-time of the electrical high speed serial link signals were measured and showed that the SPEC7 was $\approx$35 ps faster than the SPEC.
This fact causes an asymmetry on the link and hence an offset in PPS residual.
For example, the PPS residual will increase in the case a \enquote{slow} transmitting SPEC leader is combined with a \enquote{fast} transmitting SPEC7 follower.
This phenomenon explains the differences in $\Delta$ for combinations F and G.

We investigated amplitude to phase modulation of the limiting amplifier in the SFP Loopback module~\cite{sfp_loopback_module} that is used during the electrical absolute calibration procedure~\cite{EE_abscal}.
Measurements showed a 4.5 ps per 100 mV\textsubscript{pp} relation.
The small difference in serializer amplitude between SPEC (700 mV\textsubscript{pp}) and SPEC7 (500 mV\textsubscript{pp}) could explain a 9 ps error.
Limiting amplifier amplitude to phase modulation effects should be taken into account for high accuracy absolute calibrated PTP-WR links.
The contribution of latency variation due to this phenomena (on serializer receivers and SFP modules) needs further study.

Unfortunately the large offsets $\mu$ for all measured combinations \enquote{A} to \enquote{G} could not yet be explained satisfactorily and are still an open question for which the authors do not have a clear answer yet.

The PPS residual measurement results for combinations E-G form a triangle which enables us to compare results with respect to each other.
SPEC\textsubscript{VTT} seems to have a positive offset of $\approx{}$100 ps w.r.t. SPEC\textsubscript{Nikhef} and SPEC7\textsubscript{Sn01\_Nikhef} which might be due to calibration at different laboratories.
Both SPEC\textsubscript{Nikhef} and SPEC\textsubscript{VTT} suffer a positive offset 
of $\approx{}$130 ps w.r.t. SPEC7\textsubscript{Sn01\_Nikhef}, probably due to their different hardware.

The accuracy needed for electrical absolute calibration and PPS residual measurements is at the limits of our measurement equipment.
The significant measurement uncertainties prohibit us from providing firm conclusions.
Currently the Time Interval Counter is the main contribution to uncertainty.
Future studies should try to avoid the two-step electrical absolute calibration measurement as described in section \ref{sec:theory}.
Uncertainty due to cable manipulation can be decreased by using high quality phase stable cables.

\section{Conclusion}

The results of the measurements are not what the authors had anticipated.
Our goal was to prove interoperability between PTP-WR devices that were electrical absolute calibrated at different laboratories.
This endeavor was successful to some extent.
Differences between the laboratories were observed (NPL$\leftrightarrow$Nikhef:  $\mu\approx{504}$ ps; VTT$\leftrightarrow$Nikhef: $\mu$ = 107 ps).
Combinations of different types of PTP-WR devices showed significant PPS residual offsets that are not yet fully understood.
The authors encourage further study to find an answer for this unsolved puzzle.

\begin{appendix}
\end{appendix}
\section*{Funding Information}
\label{sec:FundingInformation}

This project 17IND14 WRITE has received funding from the EMPIR programme co-financed by the Participating States and from the European Union’s Horizon 2020 research and innovation programme.
\section*{Acknowledgement}
\label{sec:Acknowledgement}
The authors thank Henk Peek and Guido Visser (Nikhef) for discussions trying to shed a light on the possible causes of the PPS residual offsets found, Pascal Bos (Nikhef) for scripting- and climate chamber measurement support.

%\section{Extended error analysis}
%This is the formal error analysis.
%It was not fully performed but the estimation for the error found by the simpler method (and used in the main text) was checked based on these equations.\\
%
%\input{supplement_extended_errors}

%%
%% Bibliography for full text
%%
\bibliographystyle{unsrt}
\bibliography{journalnames,literature}

\addcontentsline{toc}{section}{References}
\begin{thebibliography}{9}
	
\bibitem{WR}
  J. Serrano, P. Alvarez, M. Cattin, E. Garcia Cota, J. Lewis, P. Moreira, T. Wlostowski, G. Gaderer, P. Loschmidt, J. Dedi\v{c}, R. B\"{a}r, T. Fleck, M. Kreider, C. Prados, and S. Rauch,
  \textit{"The White Rabbit project"}, in Proceedings of ICALEPCS2009, Kobe, Japan, (2009), pp. 93–95.

\bibitem{IEEE1588-2019}
  IEEE Standards Association,
  \textit{"IEEE Standard for a Precision Clock Synchronization Protocol for Networked Measurement and Control Systems, IEEE Std 1588-2019, (2019)."}, May 28 2018.
  [On-line]:    
  \url{https://standards.ieee.org/standard/1588-2019.html} [Oct. 19, 2020]

\bibitem{WRITE}
  \textit{"White Rabbit for Industrial Timing Enhancement"}, EMPIR 17IND14.
  [On-line]:
  \url{http://empir.npl.co.uk/write/} [Oct. 19, 2020]

\bibitem{Abscal_ISPCS_2018}
  H. Z. Peek, P. P. M. Jansweijer,
  \textit{"White Rabbit Absolute Calibration, ISPCS 2018, (2018)."}, 30 Sept.-5 Oct. 2018.
  [On-line]:    
  \url{https://doi.org/10.1109/ISPCS.2018.8543067} [Oct. 19, 2020]

\bibitem{EE_abscal}
  P. P. M. Jansweijer, H. Z. Peek, T. J. Pinkert, G. C. Visser,
  \textit{"White Rabbit Electrical Absolute Calibration Procedure"}, Version 1.1. August 28, 2018.
  [On-line]:
  \url{https://ohwr.org/project/wr-calibration/wikis/Electrical-absolute-calibration} [Oct. 19, 2020]

\bibitem{SPEC}
  OHWR.org
  \textit{"Simple PCIe FMC carrier (SPEC)"}.
  [On-line]:
  \url{https://ohwr.org/project/spec/wikis/home} [Oct. 20, 2020]

\bibitem{SPEC7}
  OHWR.org
  \textit{"Simple PCIe FMC carrier 7 (SPEC7)"}.
  [On-line]:
  \url{https://ohwr.org/project/spec7/wikis/home} [Oct. 20, 2020]

%\bibitem{CLBv3}
%  KM3NeT
%  \textit{"Central Logic Board (CLBv3) of a Digital Optical Module"} (version 3, based on Artix-7).
%  [On-line]: \url{http://www.km3net.org/} [Oct. 28, 2020]
%
\bibitem{Nikhef}
  \textit{"National Institute for Subatomic Physics (Nikhef)"}, Amsterdam, The Netherlands
  [On-line]:
  \url{https://www.nikhef.nl/en/} [Oct. 20, 2020]
    
\bibitem{VTT}
  \textit{"VTT MIKES"}, Helsinki, Finland
  [On-line]:
  \url{https://www.vtt.fi/} [Oct. 20, 2020]

\bibitem{NPL}
  \textit{"National Physical Laboratory (NPL)"}, London, United Kingdom
  [On-line]:
  \url{https://www.npl.co.uk/} [Oct. 20, 2020]

\bibitem{DIO}
  OHWR.org
  \textit{"fmc-dio-5chttla FMC 5-channel Digital I/O module"}.
  [On-line]:
  \url{https://ohwr.org/project/fmc-dio-5chttla/wikis/home} [Oct. 20, 2020]

\bibitem{sfp_loopback_module}
  OHWR.org
  \textit{"SFP+ timing calibration module"}.
  [On-line]:
  \url{https://ohwr.org/project/wr-calibration/wikis/uploads/22e40e135b6878aeb45c0af53b61e900/Data_Sheet_SFP_Calibration_Module.pdf} [Nov. 4, 2020]
    
\bibitem{DAC_report}
  N. Boukadida, P. Jansweijer,
  \textit{SFP Direct Attach Copper (DAC) cable electrical delay measurement to facilitate the electrical absolute calibration interoperability test for the WRITE project}.
  [on-line]:
  \url{https://ohwr.org/project/write/wikis/Calibration%20inter-comparison%20test}

\bibitem{wr-cores}
  OHWR.org
  \textit{White Rabbit PTP Core (WRPC)}.
  [on-line]:
  \url{https://ohwr.org/project/wr-cores/wikis/wrpc-core}

\begin{comment}
\bibitem{WR_Clocks}
  Rizzi, M., Lipiński, M., Ferrari, P., Rinaldi, S., Flammini, A.
  \textit{White Rabbit clock synchronization: ultimate limits on close-in phase noise and short-term stability due to FPGA implementation}.
  [on-line]:
  \url{http://dx.doi.org/10.1109/TUFFC.2018.2851842}

\bibitem{Meas_paper}
  H. Z. Peek, T. J. Pinkert, P. P. M. Jansweijer, J. C. J. Koelemeij,
  \textit{"Measurement of optical to electrical andelectrical to optical delays with ps-leveluncertainty"}, May 28 2018.
  [On-line]:    
  \url{https://www.osapublishing.org/DirectPDFAccess/D912487B-9C43-7F82-D2037CEE41D0AA95_389934/oe-26-11-14650.pdf?da=1&id=389934&seq=0&mobile=no} [Mar. 04, 2020]

\bibitem{SFP-crate}
  P. Jansweijer,
  \textit{"SFP+ I2C"}, Apr. 10 2019.
  [On-line]:   
  \url{https://www.ohwr.org/project/sfp-plus-i2c/wikis/home} [Mar. 04, 2020]

\bibitem{Cal_plan}
  P. Jansweijer,
  \textit{"Calibration procedure for calibrating the SFP Timing Calibration Modules"}, Mar. 05 2019.
  [On-line]:    
  \url{https://www.ohwr.org/project/wr-calibration/wikis/uploads/6917301c07ccfe912e50f06cdd24b111/SFP_TimCalMod_CalProcedure.pdf} [Mar. 04, 2020]
\end{comment}



\end{thebibliography}

\newpage

\end{document}